# Tackling dataset curation challenges towards reliable machine learning: a case study on thermoelectric materials

Shoeb Athar, Adrien Mecibah, and Philippe Jund*

ICGM, Univ Montpellier, CNRS, ENSCM, 34293 Montpellier, France

*Corresponding author. E-mail address: philippe.jund@umontpellier.fr (Philippe JUND)

**Abstract:** Machine Learning (ML) driven discovery of novel and efficient thermoelectric (TE) materials warrants experimental TE datasets of high volume, diversity, and quality. While the largest publicly available dataset, Starrydata2, has a high data volume, it contains inaccurate data due to the inherent limitations of Large Language Model (LLM)-assisted data curation, ambiguous nomenclature and complex formulas of materials in the literature. Another unaddressed issue is the inclusion of multi-source experimental data, with high standard deviations and without synthesis information. Using half-Heusler (hH) materials as an example, this work is aimed at first highlighting these errors and inconsistencies which cannot be filtered with conventional dataset curation workflows. We then propose a statistical round-robin error-based data filtering method to address these issues, a method that can be applied to filter any other material property. Lastly, a hybrid dataset creation workflow, involving data from Starrydata2 and manual extraction, is proposed and the resulting dataset is analyzed and compared against Starrydata2.

## 1. Introduction

Fossil fuel depletion and increased global power consumption are driving factors in the hunt for green energy alternatives and advanced energy recovery technologies for effective energy usage. As almost 60% of used fossil energy is lost as waste heat, finding highly efficient thermoelectric materials which produce electricity directly from heat employing the Seebeck effect is highly relevant. The performance of a TE material is rated by the dimensionless Figure of merit $zT$ given by [1] $zT = \frac{S^2\sigma}{\kappa}T$ (where $S$ is the Seebeck coefficient (V.K$^{-1}$), $T$ is the working temperature (K), $\sigma$ is the electrical conductivity (S.m$^{-1}$), and $\kappa$ is the thermal conductivity (W.m$^{-1}$K$^{-1}$). A good TE material has both high electrical conductivity and Seebeck coefficient – coupled together as $S^2\sigma$ in the TE power factor (PF) – while $\kappa$ should be minimal. Identifying such materials is highly challenging due to the intricate interplay between these properties. There has been a tremendous surge in research on mid to high-temperature TE materials over the last decades, resulting in a range of high-performance materials [2]. Yet, this has not resulted in large scale applications due to several factors involving, but not limited to, presence of poisonous, expensive and scarce elements such as Pb, Te, Ge, etc.; mechanical instability of TE materials and modules under thermal and mechanical loads; etc. [3-5]. Machine Learning can facilitate an accelerated discovery of novel high-performance TE materials through an exhaustive search of vast compositional spaces. However, since an ML model is as good as its training data, the generalizability of even state-of-the-art ML algorithms for TE materials discovery remains doubtful - as corroborated by the lack of experimental validation [6]- due to the scarcity of large curated TE datasets.

Machine Learning requires data and several research groups have attempted to build extensive datasets of TE materials containing the different TE properties. These datasets are either based on theoretical calculations, such as DFT, or from experimental data generated *in-house* or reported in the literature. Among the DFT-based datasets, the one developed by Ricci et al. (2017) [7] is the largest dataset containing the Seebeck coefficient and electrical conductivity values of 48000 compositions at different temperatures calculated with the Boltztrap code [8]. The JARVIS-DFT database [9] also has an impressive record of Boltztrap-calculated S, σ, and PF values of 22190 compositions [10]. While these large databases can be readily used for ML, a comprehensive TE characterization of a material warrants the inclusion of all TE properties making up the TE figure-of-merit *zT*. Nevertheless, the absence of thermal conductivity values in these datasets, (particularly lattice thermal conductivity) renders this task infeasible. The other DFT-based datasets, by Jaafreh et al (2021) [11], Tranas et al. (2021) [12], and Bhattacharjee et al. (2022) [13], containing $\kappa_{lat}$ entries lack the electronic part of *zT* and are much smaller. This may be attributable to the relatively higher cost of DFT calculations of $\kappa_{lat}$ against electronic properties. This cost becomes further prohibitive when dealing with doped compounds since the supercells needed to calculate the phonon spectrum become huge and not easy to handle in *ab initio* simulations.

Moreover, an exhaustive high-throughput screening of a material class, half-Heusler or otherwise, with all possible dopants and concentrations, therefore, requires an experimental dataset containing all TE properties of experimentally synthesized pure and doped materials. Gaultois et al. (2013) [14] were the first to publish a manually extracted (plotdigitizer) [15] experimental TE dataset (of *currently* ~500 materials) containing all TE properties: *zT*, S, σ, κ. With the advent of Large Language Models (LLMs) based data-mining and data-curation, larger experimental TE datasets, such as ChemDataExtractor [16], Starrydata2 [17, 18], and GPTArticleExtractor [19], have evolved lately. Among the experimental datasets, Starrydata2, is one of the largest and the most widely used publicly available TE materials datasets curated by developing an automated retrieval system using commercial Large Language Models (LLMs) [20]. It hosts the TE properties of approximately 50,000 physical samples obtained under different temperatures from around 10,000 publications including more than 130,000 curve data relevant to various TE systems (e.g., Zintl and half-Heusler compounds) [20, 21].

Notably, though adequate focus has been directed towards optimization of the ML algorithms, there are ambiguities and inadequacies in addressing the data preprocessing challenges particularly associated with the multi-source experimental data and inaccurate data embedded in publicly available datasets like Starrydata2 - which is the source for several recent ML-related studies on TE materials [21-24]. Starrydata2, though an invaluable resource, incorporates a significant degree of noise [23], and in fact inconsistent or inaccurate data, originating from data extraction for materials obtained via non-conventional synthesis routes; ambiguous nomenclature and complex formulas of materials in the literature; and inherent limitations of LLM-assisted data curation, etc.. This includes multi-source data for same composition; wrong compositions; non-uniform temperature dependencies of TE properties for certain materials resulting in asymmetric up-sampling (Section S1 and Figure S1 in the supplementary information (SI)), negative *zT* values, etc.

It is important to note that while there can be no machine learning without data, there can be no accurate ML without "good" data. The aim of this work is, therefore, not to challenge Starrydata2 but to highlight and circumvent the inherent flaws of a database curated with LLM, particularly, the challenges associated with multi-source and wrong compositions, which go beyond the regular definition of noisy data, which warrant significant attention.

## 2. Results and Discussion

**2.1 Analysis of Starrydata2**

**2.1.1 Multiple-source data:** Starrydata2 database is built by extraction of numeric data from the curves in the plot images of TE properties ($S$, $\sigma$, $\kappa$, and $zT$) published in the literature. All the data entries are assigned a unique sample and Figure ID as well as the DOI of the publication they were extracted from. A quick query for the curve entries for TiNiSn (half-Heusler), for example, renders the curves set shown in Figure 1. Intuitively, these differences in $zT$ values can be attributed to the synthesis conditions. The usual synthesis route for half-Heusler materials is arc or levitation melting followed by annealing (for ~one week) to produce homogeneity and improve the sample's atomic ordering; then an optional nano-structuring (ball-milling) step, and finally a densification process by hot-pressing (HP) or spark plasma sintering (SPS) [25]. Upon further investigation for the two visible outliers, with DOI: 10.1016/j.actamat.2015.11.022 ($zT_{max} \approx 0.978$) from [26] and DOI: 10.3390/ma11020262 ($zT_{max} \approx 0.195$) from [27], the high $zT$ values for the former is attributable to the highly optimized synthesis conditions, and the low $zT$ values for the latter to the absence of any densification process in the usual synthesis route. However, this information is not reported in Starrydata2 for most of the DOIs [22]. In fact, there are several entries for materials that were synthesized using non-conventional routes e.g. sputtering, melt spinning, high-pressure torsion, no densification, but reported without synthesis information. In addition to this, even for a given TE material, the TE properties can have inter-lab measurement uncertainty errors, called round-robin errors, between ±11.5 to ±19%. [28, 29].

A common tactic to deal with this large spread of $zT$ curves is to take the average of all the reported $zT$ values for a given composition [21, 24]. However, this becomes problematic for non-normal distributions since the arithmetic average values will be skewed from the mode (most consistently reported values), especially, when the standard deviation is large. Furthermore, if the composition in question is pure, and there are $zT$ values reported for the doped counterparts in some DOIs, the average $zT$ values for the pure composition (calculated from all DOIs) may be (almost) equal or even larger to those of doped compositions reported in other DOIs. When the average **$zT$** value happens to be almost equal to the doped compositions, this will nullify the effect of dopants, and in case, it happens to be even larger, it will attribute a detrimental doping effect. In a nutshell, taking average values of pure compositions would deprive us from the opportunity to convey the realistic effect of doping on the **$zT$** (or other TE properties) values, and defeats the purpose of ML. Starrydata2 has around 1823 compositions with multiple curves, as provided in the supplementary material (SM), suggesting the enormity of the problem.

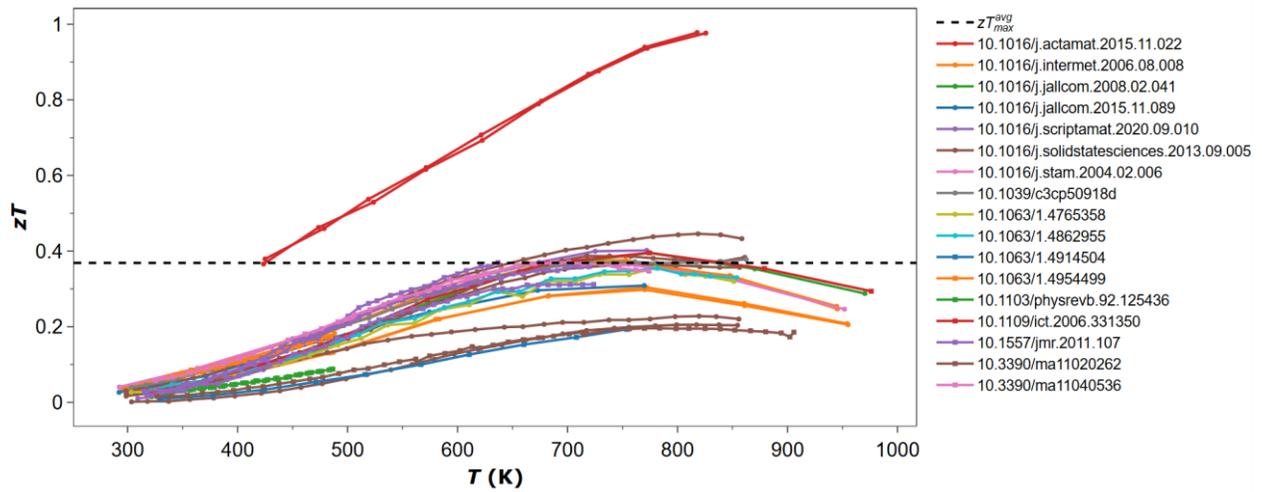

Figure 1: $zT$ vs. T curves for TiNiSn in Starrydata2.

**2.1.2 Wrong compositions:** LLM-assisted data curation have its limits especially when dealing with ambiguous naming standards of materials. As a result, Starrydata2 too has entries with wrong compositions. Here we discuss two instances of NbFeSb and TiCoSb hH materials, as shown in Figures 2 and 3, which aptly sums up the limitation of LLM in data curation. A query of the reported $zT$ vs. T curves for the NbFeSb composition renders the curves set shown in Figure 2 (a) on a log scale. Clearly, the $zT$ values, in the curves of samples 15630, 15631, and 15632 respectively, are way higher than for the other curves. Investigating the source DOI: 10.1002/advs.201800278 [30] reveals that these curves, originate from the engineering **$ZT$ ($ZT_{eng}$)** and average **$ZT$ ($ZT_{avg}$)** values reported for $Nb_{0.95}Hf_{0.05}FeSb$ as shown in Figure 2 (b) and 2 (c). Both $ZT_{eng}$ and $ZT_{avg}$ are different quantities than $zT$. The $ZT_{avg}$ is the **s**imple arithmetic mean of the instantaneous $zT$ over a given temperature range, whereas $ZT_{eng}$ is a recently defined figure-of-merit [31] that accounts for the operating temperature difference ($\Delta T$) to predict the practical performance of a TE material and is expressed as:

$$ZT_{\text{eng}} = \frac{(\int_{T_C}^{T_h} S(T)dT)^2 \int_{T_C}^{T_h} \sigma(T)dT}{\int_{T_C}^{T_h} \kappa(T)dT} \Delta T \qquad \text{Equation 1}$$

In these plots, the authors wanted to compare $ZT_{\text{eng}}$ and $ZT_{\text{avg}}$ of $Nb_{0.95}Hf_{0.05}FeSb$ (labelled as **"this work"**) against the state-of-the art p-type doped-NbFeSb ($Nb_{0.95}Ti_{0.05}FeSb$ and $Nb_{0.88}Hf_{0.12}FeSb$) materials from other works labelled as **"ref [33]"** [32] and **"ref [51]"** [33]. In the published plot images, they preferred to refer to the materials with the name of their host system (NbFeSb) instead of the actual compositions. Consequently, Starrydata2 reported these $ZT_{\text{eng}}/ZT_{\text{avg}}$ vs T curves as experimental $zT$ vs. T curves for the pure NbFeSb compositions. Since these errors originate from non-standard naming, it will be unfair to square the blame on LLM assistance for data curation here. However, there are other instances where the source of error is purely the LLM-assisted data curation.

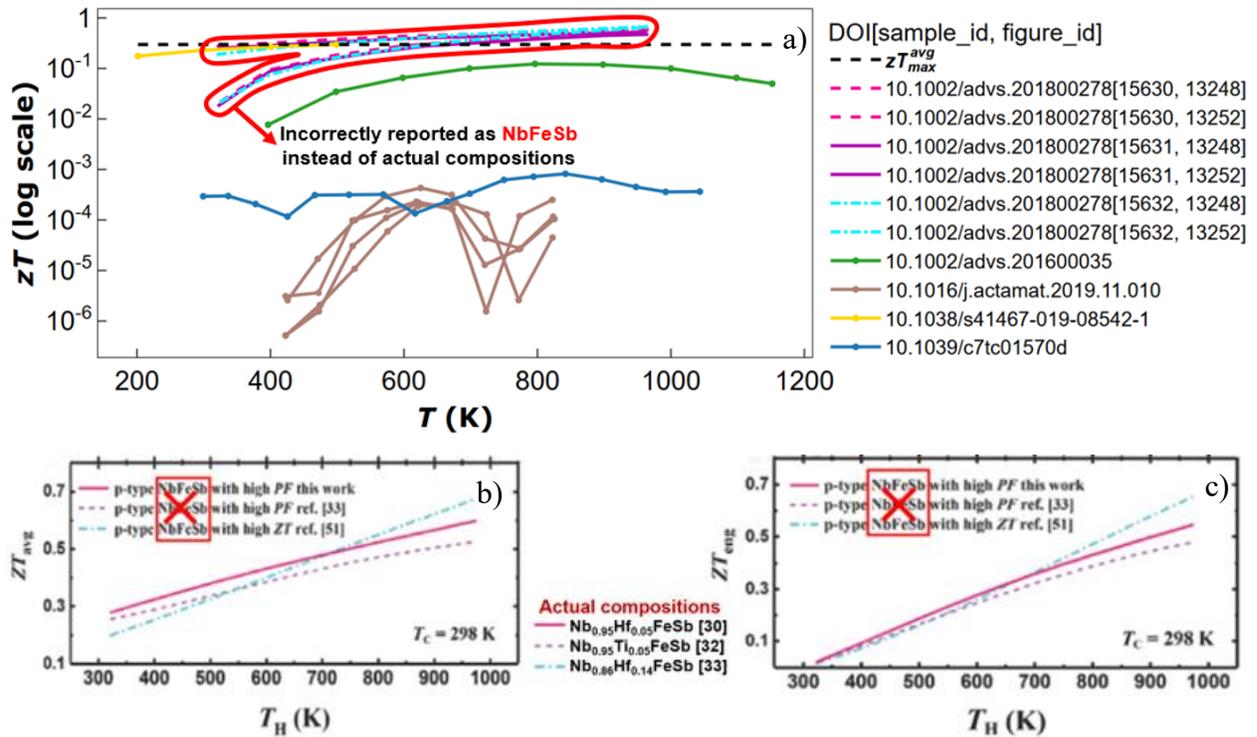

Figure 2: a) $zT$ vs. $T$ curves for NbFeSb reported in Starrydata2; b) $ZT_{avg}$ vs. T and c) $ZT_{eng}$ vs. T curves for doped NbFeSb based half-Heusler materials as published by [30] (actual compositions from original references shown in the middle space).

Let us now consider the case of TiCoSb. Figure 3 (a) lists $zT$ vs. T curves for all reported entries for this composition on a log scale. All the curves from the DOI: 10.1021/cm3011343 [34], show two $zT_{max}$ value ranges, one >0.3 and the other <0.04. Interestingly, no other DOI reports $zT_{max}$ value above 0.3 for TiCoSb. Investigating the source DOI reveals that in this work the authors have reported the $zT$ vs. T values of TiNiSn and TiCoSb using special microwave (MW) assisted synthesis to compare against the conventional arc-melting (AM) synthesis route. The source plots, and the curves therein, for these two materials, as shown in Figure 3 (a) and 3 (b), are inaccurately considered by Starrydata2 to be belonging to only TiCoSb compositions. This resulted in $zT$ vs. T curves (assigned same sample ID) with large differences in the $zT_{max}$ (or $zT$ values) for the same composition. In both the cases of NbFeSb and TiCoSb, the resulting average $zT_{max}$ (or average $zT$) values have been inflated due to wrong curves. Needless to say that these are just selected examples for a single class of half-Heusler materials, and one may expect a large number of such errors across all materials classes given the size of Starrydata2.

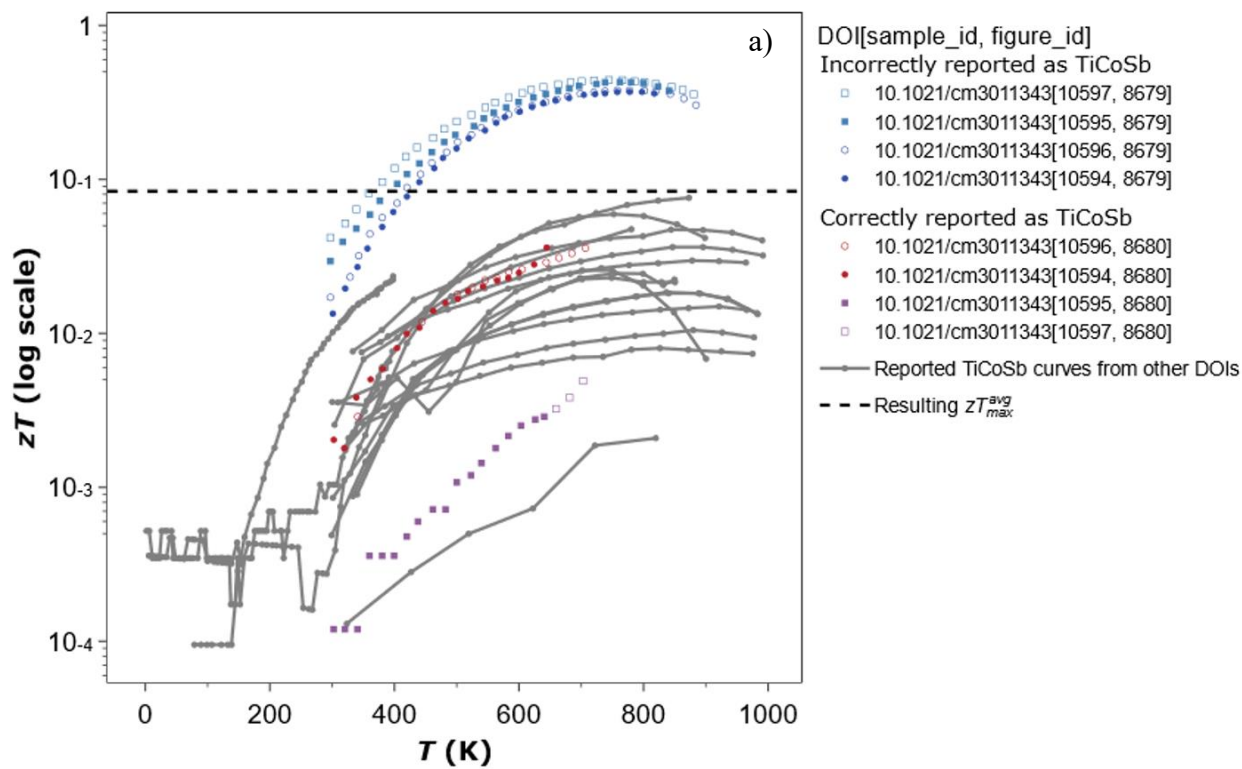

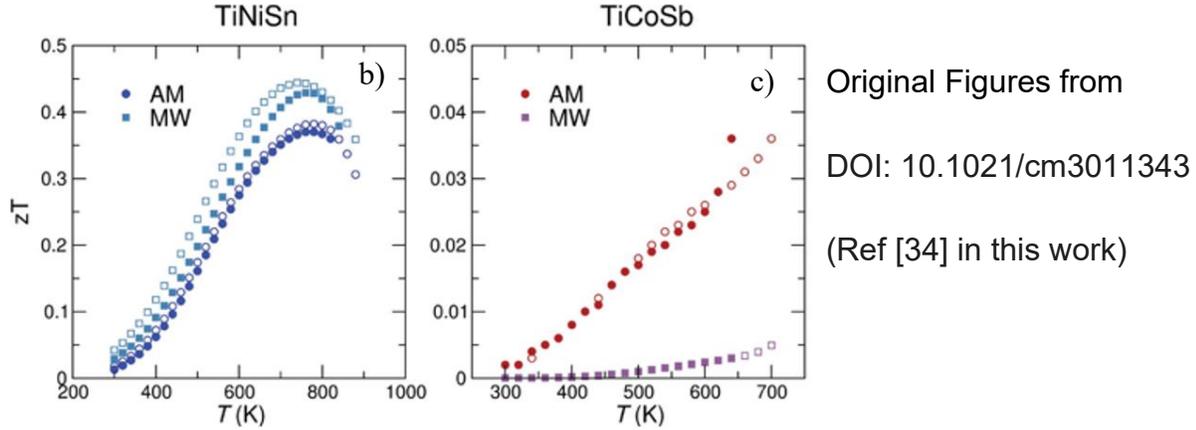

Figure 3: a) $zT$ vs. T curves reported for TiCoSb in Starrydata2; $zT$ vs. T curves for b) TiNiSn and c) TiCoSb published in ref [34].

In this work, using the case of TiNiSn based hH materials from Starrydata2, we propose a statistical bin-filtering method based on round-robin errors to screen the multi-source $zT$ vs. T curves reported for the same compositions. With the examples of NbFeSb and TiCoSb, we also demonstrate how this method can be used to address the issue of wrong compositions, incorporated in -but not limited to- Starrydata2 due to ambiguous nomenclature and limitations of LLM-assisted data curation. Finally, combining data from existing literature and Starrydata2 for theTE hH materials, we propose a hybrid dataset curation workflow to create a manually curated dataset - of hierarchical structure - comprising materials, compositions, and temperature dependencies (as illustrated in Table S1 in the SI).

## 2.2 Round-robin bin filtering method

We propose a filtering method based on the round-robin error of ±15% (rounded-off) reported for half-Heusler materials [28], to address the issue of multi-source data and wrong compositions. As evident from Figure 1, the majority of the reported curves tend to center around a $zT_{max}$ value between 0.3 and 0.4. We start with the underlying assumption that the average $zT_{max}$ ($zT_{max}^{avg}$) is indeed the true $zT_{max}$ of this material, as has been the strategy so far. The method, as illustrated in Figure 4, can then be explained in the following steps as follows:

Step 1: We extract all the reported $zT$ vs. T curves for a given composition (TiNiSn in this example) and collate their $zT_{max}$ values.

Step 2: We define a bin of width equal to the round-robin error of ±15% of the average $zT_{max}$ ($zT_{max}^{avg}$) value of all the curves with $zT_{max}^{avg}$ at the center. This bin, whilst maintaining its width, can then be multiplied to cover the entire range of $zT_{max}$ values reported for the material. The next step is to fill these bins with the DOI (without repetitions) whose $zT_{max}$ values fall within that bin. Now, if our initial assumption of the $zT_{max}^{avg}$ is true, the majority of the curves shall populate the bin around this value. Else, however, if the $zT_{max}^{avg}$ value is inflated, due to large differences in $zT$ values for different DOI entries for the same composition in the dataset (as for the NbFeSb and TiCoSb curves shown in Figures 2 and 3), one would expect other bins to be more populated based on the most consistently reported $zT_{max}$ values. In any case, we proceed by shortlisting the bin most populated with DOIs, designated as « selected bin », for data extraction (in case if two or more bins have same number of DOIs, we select the one which has more $zT$ vs. T curves).

Step 3: For selecting a unique DOI for the pure material, we now adopt an approach based on information gain, and select the DOI with the most reported doped materials. If two or more DOIs have same number of doped materials, we select the one with most reported doped compositions.

Step 4: For some of the materials, there are multiple curves reported for the same pure composition. This can be a consequence of synthesis optimization, for example in the case of TiNiSn, where properties of multiple samples are reported for different synthesis steps. Alternatively, it can result from wrong data extraction, as discussed with TiCoSb. Therefore, building upon the idea explained in step 2, we now calculate the $zT_{max}^{avg}$ of the **selected bin** and select the most representative curve, from the selected DOI, based on the $zT_{max}$ value closest to this $zT_{max}^{avg}$ value. All the curves of doped materials/compositions, corresponding to this selected curve for pure material, are also selected.

Step 5: Once the DOI and curve have been selected, we can still consider other candidate DOIs/curves, discarded in step 2 and 3, for additional doped materials. However, these candidate DOIs must report additional but different doped materials than those already reported in the selected DOI and the $zT_{max}$ of the pure material in these candidate (additional) DOIs must fall within ±15% (round-robin error) of $zT_{max}$ reported for the pure material in the selected curve (see Figure 4).

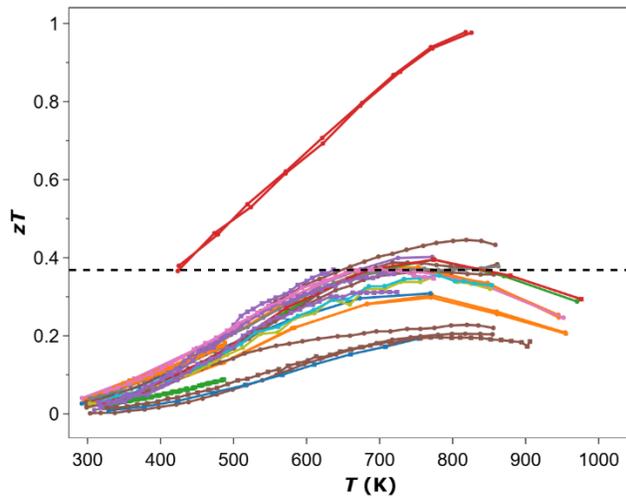
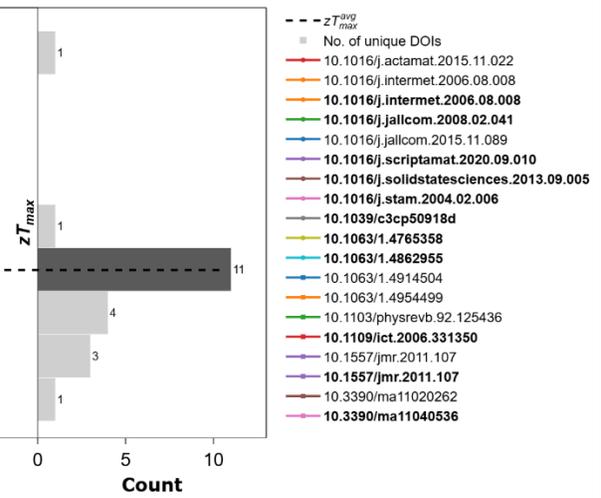
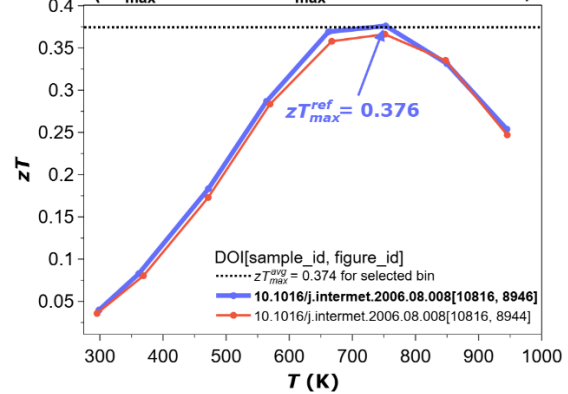
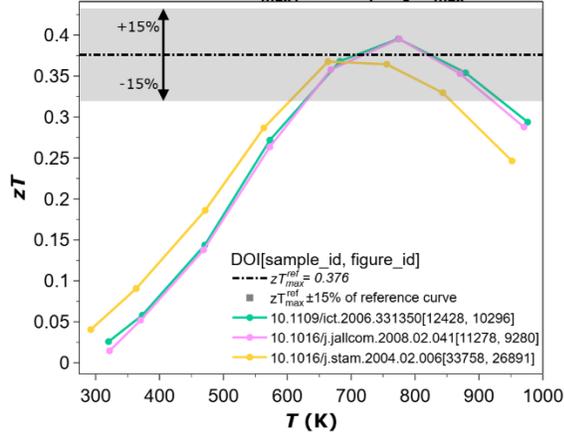
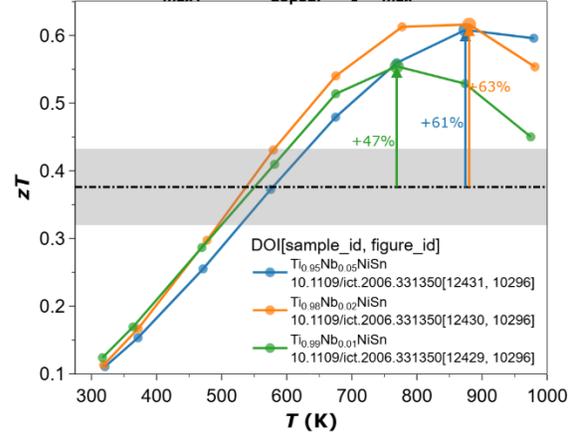

Figure 4: Workflow of round-robin bin filtering method.

Step 6: We finally select curves for doped materials from candidate DOIs if the doped materials reported in these works have a $zT_{max}$ increment/decrement greater than 15% against the pure material reported in the selected curve. This is to ensure that the positive or negative effect of doping is captured without having to select multiple curves for the pure material in question.

It is still possible to have some data loss due to filtering, but this approach strikes a fine balance between data quality and data volume. This method, in addition, can address the issue of wrong compositions/inaccurate data. Figure 2 and 3 show that the curves for the wrong compositions distinguish themselves from the curves of the accurately reported compositions. Isolating these outliers (through binning $zT_{max}$), which is exactly what this method accomplishes, from the majority of accurate curves can suffice to filter the wrong data. This is shown in Figure 5 for NbFeSb and TiCoSb, respectively. The GitLab link to access the script is provide in the SM. This script can be adapted to any TE property, structural prototype or composition, and dataset. In fact, it can be applied to filter any material composition-based property by applying, a user-defined domain specific interlaboratory measurement uncertainty range error (e.g. round-robin errors for TE materials).

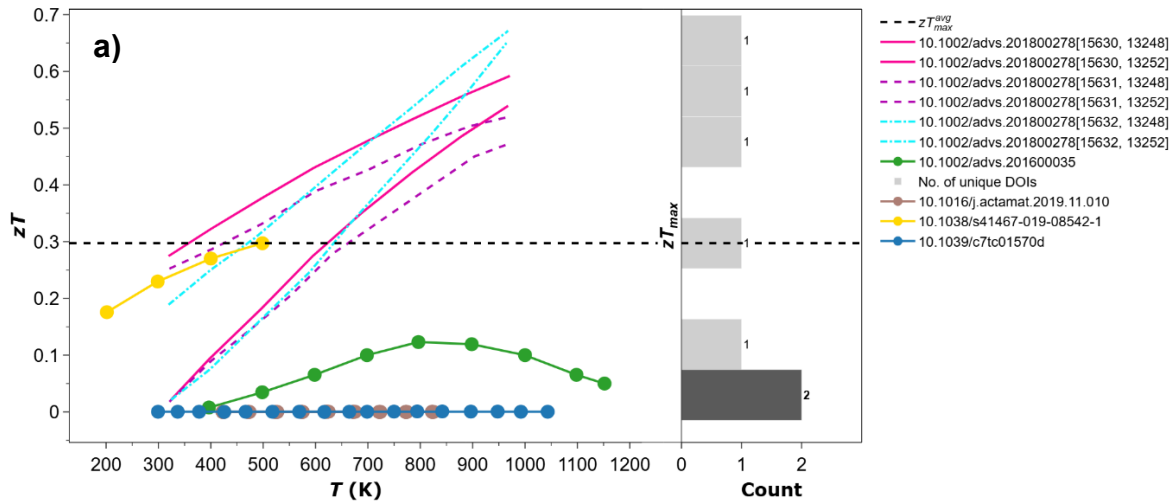

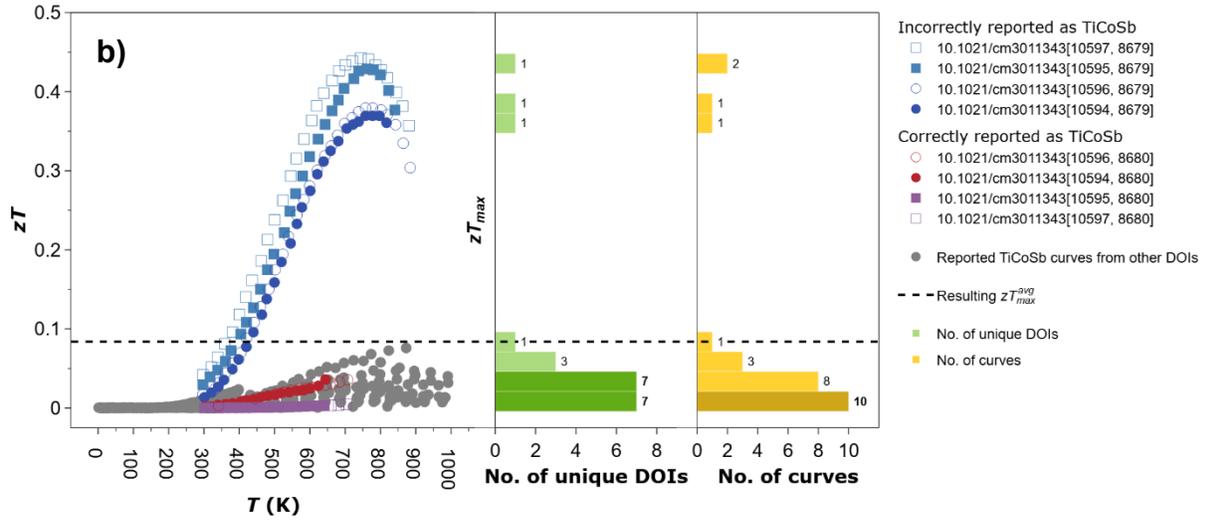

Figure 5: $zT$ vs T curves reported for a) NbFeSb curves and b) TiCoSb curves in Starrydata2 with the histogram for $zT_{max}$ values.

## 2.3 Dataset curation workflow

Due to the challenges associated with data-mining, as discussed before, manual data extraction from the literature still remains the mainstream method for small datasets dependent material science properties [35]. In this work we propose a hybrid approach that involves manual data extraction from the literature as well as data incorporation from existing datasets (Starrydata2) - datasets (Starrydata2) - using the data filtering technique shown in Figure 4 incorporated in the workflow shown in Figure 6. Foremost, we manually extracted the $zT$ vs. T curves of stoichiometric half-Heusler materials, of type $A_{1-x}As_xB_{1-y}Bs_yC_{1-z}Cs_z$ (where A, B: transition metals, C: p-block elements as in [36]) and As, Bs, and Cs are substituents at the A, B and C sites with x, y, and z as their respective concentrations), from published articles using a free software, PlotDigitizer [15], in addition to the curves reported in Starrydata2.

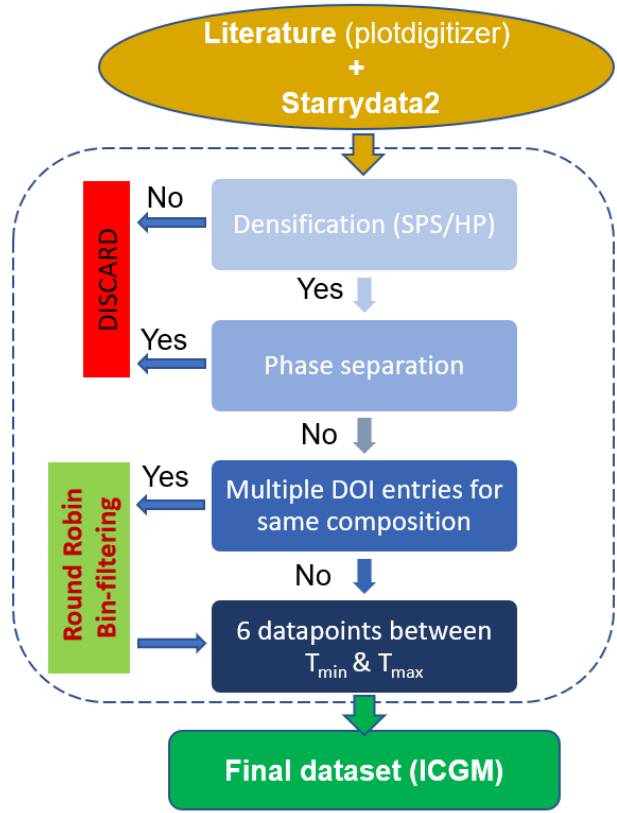

Figure 6: Dataset creation/curation workflow.

A total of 192 DOIs were vetted through our filtering process which first employs the exclusion of non-densified samples i.e. inclusion of only hot-pressed (HP) or spark plasma sintered (SPS) samples. Then, phase engineered materials or multi-phasic materials are dropped. Then, to address the issue of multi-source experimental data (and possibly wrong compositions) with multiple DOIs/Curve IDs, we use the round-robin bin filtering method as discussed previously. In the last step, to avoid over-representation of some materials through higher temperature dependencies, a maximum of six experimental $zT$ vs T values were selected for each curve between the reported $T_{min}$ and $T_{max}$. Although some curves in Starrydata2 had fewer dependencies than six, no interpolations or extrapolations of $zT$ vs T values were done to up-sample or down-sample the data. Henceforth, we refer to our dataset as ICGM. The list of hH materials and compositions along with their DOIs are provided in the SM. Our dataset curation workflow and, thereby, ICGM dataset, offers the following advantages:

1. Manually curated for unique compositions without duplications

2. All densified bulk samples (SPS/HP)
3. No material from unusual synthesis processes (e.g. sputtering, melt-spinning)
4. No phase-separated materials
5. Uniform temperature dependencies for all compositions
6. Consists of all filtered entries from starrydata2 and manual literature extraction

## 2.4 Visualization and comparison of ICGM and Starrydata2

To highlight the impact of our workflow on dataset creation/curation, it is important to visualize and compare the characteristics of our dataset (ICGM) and Starrydata2 - for hH materials. The data visualization is shown in Figures 7-9. The heat map representing the elements present in the materials in both the datasets is shown in Figure 7. Sn and Sb are undoubtedly the most common elements in half-Heusler TE materials, encountered in the literature, followed by Co, Ni, Zr, Ti, Hf, Nb, etc. These facts are well established by both the datasets.

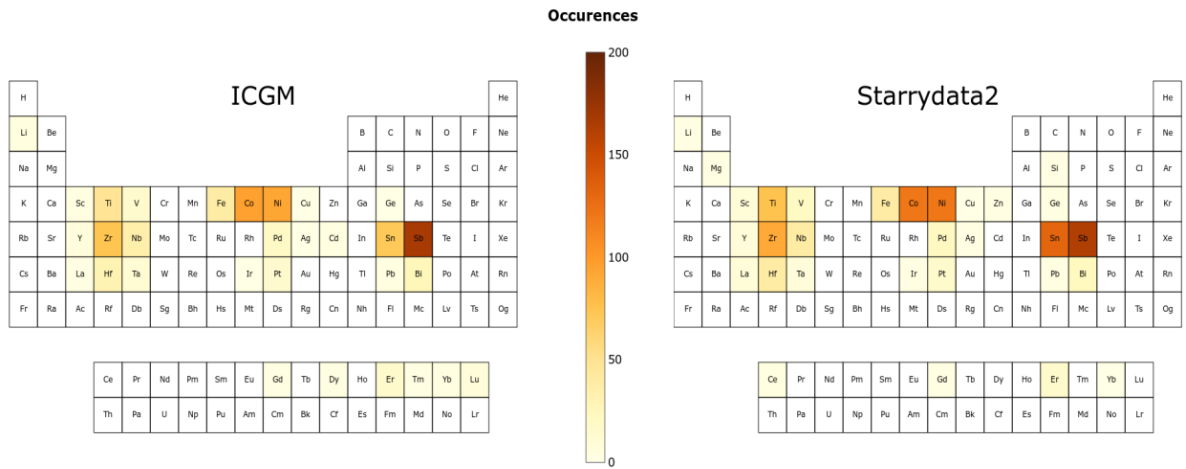

Figure 7: Heat map representation of elements in the ICGM and Starrydata2 datasets.

The only elements ICGM lacks are Al and Si thanks to the filtering of the LiAlSi material due to the absence of densification. The elements which are additionally present in ICGM are rare-earths: Dy, Tm, and Lu originating from DyNiSb, TmNiSb and LuNiSb-based materials. Here, we have shown that despite the strict filtering process the diversity of elemental representation is preserved, and in fact, enhanced using the hybrid dataset creation process (Figure 6). Figure 8 shows the $zT$ distribution of the data points in ICGM and Starrydata2 which is typically a non-normal distribution with $zT$s concentrated at lower values. This is not only archetypal for hH materials, but also for TE materials in general reported in Staryydata2 [21] and emphasizes the scarcity of high-$zT$ TE materials. Importantly, for this work, materials in ICGM are well representative of all the $zT$ values encountered for hH materials.

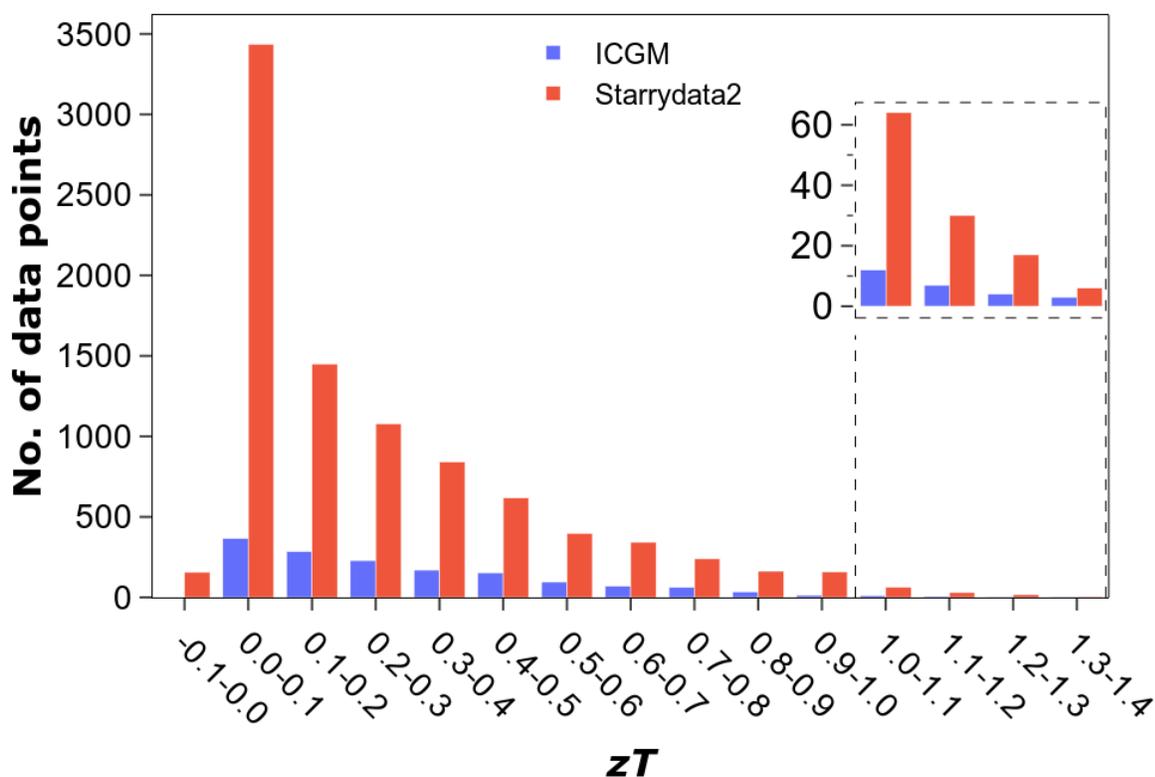

Figure 8: $zT$ distribution of the data points in ICGM and Starrydata2.

The Uniform Manifold Approximation and Projection (UMAP) [37] is a dimensionality reduction technique that allows visualization of high dimensional complex datasets whilst preserving their underlying structure. The UMAP plot of the hH materials in the two datasets is shown in Figure 9 (a). It is clear that the materials in the ICGM dataset (blue + green) effectively covers nearly a similar UMAP area compared to the area covered by Starrydata2 (green + red), and therefore is equivalent in data diversity with respect to the chemical space. Moreover, there are several materials (red) that have been discarded following the dataset curation process because they were not suitable for inclusion due to unusual synthesis, phase separation, etc. This results in a smaller number of materials and compositions in ICGM (108 and 256) compared to Starrydata2 (132 and 267) as listed in Table 1. This difference is the exact trade-off between data volume and data accuracy - albeit without a compromise on data diversity. The temperature dependencies (data points) are unsurprisingly six times higher in Starrydata2 since we did not interpolate/extrapolate them in ours as this would not have increased the data diversity.

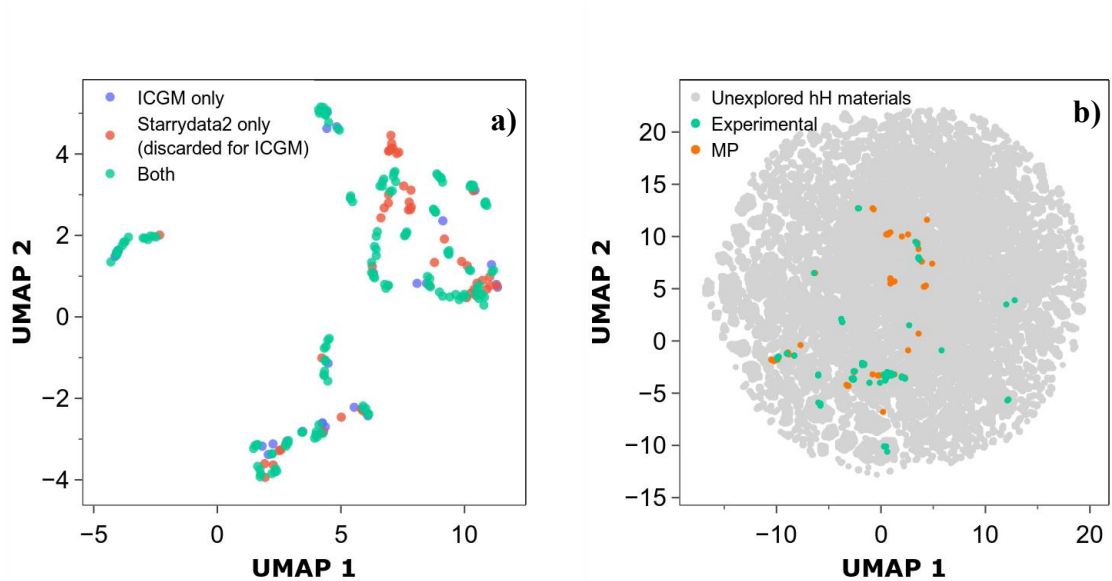

Figure 9: UMAP plots of a) half-Heusler materials in ICGM and Starrydat2 datasets and b) half-Heusler materials compared to experimental TE hHs and predicted hHs structures till date.

Table 1: Overview of data on half-Heusler materials in ICGM and Starrydata2.

| Dataset | No. of materials | No. of compositions | No. of data points |
|---|---|---|---|
| ICGM | 108 | 256 | 1521 |
| Starrydata2 | 132 | 367 | 9003 |

Lastly, we compare the combinatorial chemical space for predicted hH systems from Materials Project (MP) [38] and unexplored hH systems (with single or double doped sites) to the existing hH TE materials through UMAP plots as shown in Figure 9 (b). A clear conclusion that can be drawn is despite the high volume of 'data points', particularly in Starrydata2, these datasets are low in diversity and don't span the representative chemical space. To qualify as 'big data', the dataset must be high in volume, diversity of chemical space, and quality [39]. This aptly proves the small-data problem in machine learning for TE materials regardless of the availability of big volume datasets. Yet, our dataset creation and curation approach, demonstrated with ICGM, tries to optimize the quality, volume, and diversity from the available data in the literature while eliminating wrong data and data redundancies. The chemical space from MP offers only a slightly better data representation and diversity of the chemical space than the experimental TE data for hHs. However, there is visibly a huge unexplored chemical space that can be screened for finding efficient TE materials. The combinatorial count of possible half-Heusler materials and compositions are 23,619,600 and 2,442,210,997,860 (50 concentration increments at each doped site). This emphasizes both the enormity of the task and the importance of data-driven discovery of TE materials. Based on unexplored sections of the compositional space further experimental works can be guided to enhance the data diversity of materials in literature/datasets, and thereby, render ML feasible for TE materials.

## 3. Conclusion

Datasets form the core of any ML-related work where their volume, quality, and diversity dictate the accuracy of the ML-models. With a surge in ML-driven discovery of TE materials, the publicly available large datasets, like Starrydata2, take the center stage. Therefore, it is important to analyze and highlight the inconsistencies and inaccuracies in these datasets to guide ML-research. This work highlighted the hitherto unaddressed issue of multi-source experimental data and wrong compositions using the example of Starrydata2. Therefore, a data filtering method - applicable to any material-composition based property- was proposed to address these issues which was then followed by a hybrid dataset curation workflow using the half-Heusler materials as the structural-prototype. The resulting ICGM dataset for half-Heusler materials will be used in our future works for training with several important machine learning techniques. We believe that with a reasonable and a multi-institutional collaborative effort, it is conceivable to create a comprehensive and curated dataset covering all structural-prototypes of TE materials. Since without accurate data, ML-models, regardless of their performance metrics on test data, will remain of questionable merit; this task should be the priority of the thermoelectric research community i.e. data curation must take precedence over works on ML-algorithm optimizations.

## 4. Computational Methods

The raw dataset was downloaded from StarryData2 website [40], accessed on 27th September 2024, and stored in a csv format. All the entries for the type $A_{1-x}As_xB_{1-y}Bs_yC_{1-z}Cs_z$ hH materials were extracted. The UMAP visualization was constructed using 19 physical properties (depicted in Table S2 in the SI) of the six constituent elements of type $A_{1-x}As_xB_{1-y}Bs_yC_{1-z}Cs_z$ hH materials creating a 114-dimensional vector to use as numerical input (we consider the doped element to be the same as the pure element if the concentration is zero). The Mendeleev python library [41] was used to extract the values of these physical properties. The UMAP plot for the unexplored hH materials was constructed using combinatorial permutations of A, B, and C site elements with possible substituents as proposed by [36] at one or two parent sites. Similarly, the UMAP plot for hH materials in Materials Project was constructed using possible dopants in hH systems reported in the Materials Project database [38] with a hull-distance < 0.05 eV/atom.

*isotopes and methods to manipulate and visualize periodic table.* GitHub. <https://github.com/lmmentel/mendeleev>.

## Supplementary Material

https://src.koda.cnrs.fr/icgm-thermoelectrics-ml/dataset-curation-by-round-robin-method

## Acknowledgements

We acknowledge the financial support from the Agence Nationale de la Recherche (ANR), France under the ANR-DFG project « CombiHeusler » (ANR-24-CE92-0053-01).

## Competing Interests

The authors declare no Competing Financial or Non-Financial Interests.